\documentclass[11pt]{article}
 
\usepackage{microtype}
\usepackage[margin=1in]{geometry}
\usepackage{graphicx}
\usepackage[utf8]{inputenc}
\usepackage[ruled,vlined,linesnumbered]{algorithm2e}
\usepackage{subcaption}
\usepackage{soul,color}
\soulregister\cite7
\soulregister\ref7
\soulregister\pageref7
\usepackage{xcolor}
\usepackage{dsfont}
\usepackage{booktabs}
\usepackage{multirow}
\usepackage{hyperref}
\usepackage[multiuser,draft]{fixme}
\FXRegisterAuthor{rg}{arg}{RG}
\FXRegisterAuthor{lv}{alv}{LV}
\fxusetheme{colorsig}

\title{Round-Hashing for Data Storage: Distributed Servers and External-Memory Tables}

\author{
Roberto Grossi\\Dipartimento di Informatica, Universit\`a di Pisa, Italy\\grossi@di.unipi.it
\and
Luca Versari\\Dipartimento di Informatica, Universit\`a di Pisa,Italy\\luca.versari@di.unipi.it
}

\newcommand{\stash}{\ensuremath\mathtt{stash}\xspace}
\newcommand{\block}{\ensuremath\mathtt{block}}
\newcommand{\init}{\ensuremath\mathtt{init}}
\newcommand{\numBuckets}{\ensuremath\mathtt{numBuckets}}
\newcommand{\findBucket}{\ensuremath\mathtt{findBucket}}
\newcommand{\newBucket}{\ensuremath\mathtt{newBucket}}
\newcommand{\freeBucket}{\ensuremath\mathtt{freeBucket}}
\newcommand{\lookup}{\ensuremath\mathtt{lookup}}
\newcommand{\ins}{\ensuremath\mathtt{insert}}
\newcommand{\del}{\ensuremath\mathtt{delete}}
\newcommand{\distribute}{\ensuremath\mathtt{distribute}}
\newcommand{\false}{\ensuremath\mathit{false}}
\newcommand{\true}{\ensuremath\mathit{true}}
\newcommand{\found}{\ensuremath\mathit{found}}

\newcommand{\erfc}{\mathrm{erfc}}
\newcommand{\pow}{\mathit{pow}}
\newcommand{\pos}{\mathit{pos}}
\newcommand{\len}{\mathit{len}}
\newcommand{\modulo}{\mathbin{\%}}

\newtheorem{theorem}{Theorem}
\newtheorem{lemma}{Lemma}

\makeatother

\begin{document}

\maketitle

\begin{abstract}
  This paper proposes round-hashing, which is suitable for data
  storage on distributed servers and for implementing external-memory
  tables in which each lookup retrieves at most a single block of
  external memory, using a stash. For data storage, round-hashing is
  like consistent hashing as it avoids a full rehashing of the keys
  when new servers are added. Experiments show that the speed to serve
  requests is tenfold or more than the state of the art. In
  distributed data storage, this guarantees better throughput for
  serving requests and, moreover, greatly reduces decision times for
  which data should move to new servers as rescanning data is much
  faster.
\end{abstract}

\section{Introduction}%
\label{sec:introduction}

Consistent hashing was invented by Karger et
al.~\cite{Karger:1997:CHR:258533.258660} for shared web caching and
highest random weight hashing (also known as rendezvous hashing) was
invented by Thaler and Ravishankar~\cite{ThalerR98} for web proxy
servers. Both hashing methods were conceived independently around the
mid 90s, and shared similar goals (with different implementations):
cached web pages are assigned to servers, so that when a server goes
down, its cached web pages are reassigned to the other servers so as to
preserve their load balancing; similarly, when a new server is added,
some cached web pages are moved to it from the others.

Consistent hashing, in its basic version, maps both web pages and
servers to the circular universe $[0\ldots 2^w-1]$, where each hash value
requires $w$ bits: each web page starts from its hash value in the
circular universe and is assigned to the server whose hash value is
clockwise met first; this can be done in $O(\log m)$ time
using a search data structure of size $O(m)$ for $m$ servers. Rendezvous
hashing, for a given web page $p$, applies hashing to the pairs
$\langle p, i \rangle$ for each server $i$, and then assigns $p$ to
the server $i=i_0$ that gives the maximum hash value among these
pairs; this is computed in $O(\log m)$ time using a tree of size
$O(m)$ as discussed by Wang and Ravishankar~\cite{WangR09}. 
Table~\ref{table:horserace} reports a summary of these bounds.
Both methods apply their rule above when a server is deleted or
added. They have been successfully exploited in the industry,
e.g. Akamai for consistent hashing, and Microsoft's cache array
routing protocol (CARP) for rendezvous hashing. Among the notable
applications, it is worth mentioning Chord~\cite{Stoica:2003:CSP} for
building distributed hash tables in peer-to-peer networks (such as
BitTorrent), and Amazon's Dynamo~\cite{DeCandiaHJKLPSVV07} for
distributed and cloud computing, with data stored in main memory for
speed on a wide set of machines.

Recently, Lamping and Veach presented jump consistent
hashing~\cite{LampingV14} at Google, observing that consistent hashing
can be tailored for data centers and data storage applications in
general. In this scenario, servers cannot disappear, as this would
mean loss of valuable data; rather, they can be added to increase
storage capacity.\footnote{Data is split into shards, where each shard
  is handled by a cluster of machines with replication, thus it is not
  acceptable for shards to disappear~\cite{LampingV14}.} As a result,
the hash values ``jump'' to higher values for the keys moved to a new
bucket; moreover, the hash values are a contiguous range
$[0\ldots m-1]$ for $m$ servers, rather than a subset of $m$ integers
from $[0\ldots 2^w-1]$.  This has a dramatic impact on the performance
of the jump consistent hash, as illustrated in~\cite{LampingV14},
observing that only balance and monotonicity should be guaranteed from
the original proposal in~\cite{Karger:1997:CHR:258533.258660}.  The
auxiliary storage is just $O(1)$, as shown in
Table~\ref{table:horserace}; average query cost is the $m$-th harmonic
number, so $O(\log m)$, with no worst case guarantee.

In this paper, we study the problem of consistent hashing for data
storage in the above scenario depicted by Lamping and Veach. For the
presentation's sake, the keys are the hash values of the web pages,
and the servers are the buckets, numbered from $0$ to $m-1$, where the
keys have to be placed. At any time, we want to support the following
operations, besides the initialization:
\begin{itemize}
\item Return the current number $m$ of buckets.
\item Given a key $u$, find its corresponding bucket number in
  $[0 \ldots m-1]$.
\item Add a new bucket having number $m$, thus the range becomes
  $[0\ldots m]$.
\item Release the last bucket $m-1$, thus the range becomes
  $[0\ldots m-2]$.\footnote{This operation is not actually mentioned
    in the original setting, 
    but it comes for free in our case. }
\end{itemize}

We observe that linear hashing, introduced by
Litwin~\cite{litwin1980linear} and Larson~\cite{larson1980linear} at
the beginning of the 80s, can be successfully employed in this
scenario, thus taking $O(\log m)$ time and $O(1)$ space, as reported
in Table~\ref{table:horserace}. We use a constant slack parameter
$s_0 \ll m$ (typically $s_0=64$ or $128$) to guarantee that the number
of keys in the most populated bucket is at most $1 + 1/s_0$ times the
number of keys in the least populated bucket.

\begin{table*}[t]
  \centering
\begin{tabular}{l|cccl}
                                                          & find bucket & space & new bucket & notes \\
  \hline
  consistent hashing~\cite{Karger:1997:CHR:258533.258660} & $O(\log m)$         & $O(m)$ & $\tilde{O}(\alpha + \log m)$ & local \\
  rendezvous hashing~\cite{ThalerR98}                     & $O(\log m)$         & $O(m)$ & $O(m \alpha)$ & distributed \\
  jump consistent hash~\cite{LampingV14}                  & $\tilde{O}(\log m)$ & $O(1)$ & $O(m \alpha)$ & distributed  \\
  linear hashing~\cite{larson1980linear,litwin1980linear} & $O(\log (m/s_0))$       & $O(1)$ & $\tilde{O}(s_0 \alpha)$ & local \\
  round-hashing (ours)                                    & $O(1)$              & $O(1)$ & $\tilde{O}(s_0 \alpha)$ & local, no division \\[2mm]
\end{tabular}
\caption{Performance of the hashing methods for $m$ buckets
  (servers). Here $s_0 \ll m$ is a constant slack parameter (typically
  $s=64$ or $128$), and $\alpha = $ (number of stored keys) $/\, m$ is
  the load factor. Although creating a new bucket moves
  $O(\alpha)$ keys on the average, each hashing method can take
  different time to decide which are the keys to move: ``local'' means
  that few other buckets scan their keys, while ``distributed'' means
  that all buckets scan  their keys in parallel to decide which ones
  have to move to the new bucket. The $\tilde{O}()$ notation indicates
  an expected cost.}%
\label{table:horserace}
\end{table*}

\begin{figure*}[t]
  \scriptsize
{\small\textsl{round} $q=0$}

\fbox{0 1 2}

\vspace*{4pt}
{\small\textsl{round} $q=1$}

0 1 2 \fbox{3 4 5}

\vspace*{4pt}
{\small\textsl{round} $q=2$}

0 1 2 \fbox{6 8 10} 3 4 5 \fbox{7 9 11}

\vspace*{4pt}
{\small\textsl{round} $q=3$}

0 1 2 \fbox{12 16 20} 6 8 10 \fbox{13 17 21} 3 4 5 \fbox{14 18 22} 7 9 11 \fbox{15 19 23}

\vspace*{5pt}
{\small\textsl{round} $q=4$}

\quad step $s = s_0 = 3$

0 1 2 24 12 16 20 25 6 8 10 26 13 17 21 27 3 4 5 28 14 18 22 29 7 9 11 30 15 19 23 31

\quad step $s = 4$

0 1 2 24 32 12 16 20 25 33 6 8 10 26 34 13 17 21 27 35 3 4 5 28 36 14 18 22 29 37 7 9 11 30 38 15 19 23 31 39

\quad step $s = 5 = 2s_0-1$

\mbox{0\,1\,2\,\fbox{24\,32\,40}\,12\,16\,20\,\fbox{25\,33\,41}\,6\,8\,10\,\fbox{26\,34\,42}\,13\,17\,21\,\fbox{27\,35\,43}\,3\,4\,5\,\fbox{28\,36\,44}\,14\,18\,22\,\fbox{29\,37\,45}\,7\,9\,11\,\fbox{30\,38\,46}\,15\,19\,23\,\fbox{31\,39\,47}}

\caption{Example of round-mapping with $s_0=3$, where the
  sequences of bucket numbers are not actually materialized by our
  algorithm. Framed chunks of $s_0$ elements represent which bucket
  numbers have been added during round~$q$.}%
  \label{fig:example-round-mapping}
\end{figure*}

\paragraph*{Our first contribution} 
In this paper we present a new hashing scheme, called
\emph{round-hashing}, which applies \emph{round-mapping} to the hashed
value, allowing us to achieve $O(1)$ time and space in the worst
case. This is a desirable feature, as otherwise hashing with no
worst-case guarantee can pose security issues, such as algorithmic
complexity attacks~\cite{Bar-YosefW07,CrosbyW03} for low-bandwidth
denial of service exploiting its worst-case behavior.

Round-hashing computes the hash value of the given key and invokes
round-mapping for this value, as described in
Section~\ref{sec:round-mapping}. Here we give a glimpse using the
example with the bucket numbers shown in
Figure~\ref{fig:example-round-mapping}, where we set $s_0 = 3$.
round-mapping does not materialize these numbers, and proceeds in rounds
$q=0,1,\ldots$ when adding new bucket numbers, where at the end of
round $q$ there are $m = 2^q s_0$ bucket numbers. For instance,
consider when round $q-1=3$ ends ($m = 8s_0$) and a new bucket number
is requested, so round $q=4$ begins. 

We observe that each round is divided into steps, where each step
$s = s_0, s_0+1, \ldots, 2 s_0 - 1$ handles the next $8 = 2^{q-1}$
requests of new bucket numbers. At the end of these steps,
$2^{q-1} s_0$ bucket numbers have been added to the $2^{q-1} s_0$ ones
inherited from round $q-1$, and thus we find the claimed
$m = 2^q s_0$ bucket numbers at the end of round $q$.

Let us go back to the first request for $q=4$. It is for block number
$m=24$, which is implicitly inserted after the first $s=s_0=3$ elements,
between 2 and 12. Next comes $m=25$, so we skip other $s=3$ elements,
inserting it between 20 and 6. In general, for each step $s$, we insert a
new bucket number by skipping the next $s$ elements from the
current position. When step $s$ completes its virtual scan, it means
that $2^{q-1}$ new block numbers have been served (8 in our example),
so we can set $s := s+1$ and repeat the scan from the beginning if $s
< 2s_0$. 

In summary, assigning a new block number is simple: skip the next
$s$ elements from the current position and insert the new bucket
number; if the end of the scan is reached, increase $s$ and restart
from the beginning if $s < 2s_0$. Any time we have a
permutation of $[0 \ldots m-1]$: what is non-trivial is that, for any
given (hash value) $u \in [0 \ldots m-1]$, round-mapping computes the
block number at position $u$ in this evolving permutation, in $O(1)$
time using $O(1)$ additional memory, without general division and
modulo operations.  We refer the reader to
Section~\ref{sec:round-mapping} for a formal description.

Finally, looking back at Table~\ref{table:horserace} when a new bucket
is created, we observe that $O(\alpha)$ keys on average are moved
from the other buckets, where $\alpha$ is the load factor, namely, the
number of stored keys divided by the number $m$ of
buckets.\footnote{We are assuming, wlog, that the buckets have all the
  same size.} However, the hashing methods take different time to
decide which are the keys to move.  Specifically, consistent hashing
has to examine the $O(\alpha)$ keys in the two neighbor servers in the
worst case, and update the data structure in $O(\log m)$
time.\footnote{For the sake of discussion, we consider the basic
  version of consistent hashing, and refer the reader
  to~\cite{Karger:1997:CHR:258533.258660,LampingV14} for the version
  with multiple hash values per server.} Rendezvous hashing requires
that each bucket scans its keys and test whether the new bucket is now
the maximum for some of them. Hence all the keys are scanned,
$O(m \alpha)$, but only $O(\alpha)$ of them are moved in total.  Jump
consistent hashing needs to perform a similar task, to see which keys
``jump'' to the new bucket. Linear hashing and our round-hashing
require to scan the keys in $s_0 = O(1)$ buckets to find the
$O(\alpha)$ ones to move. Note that the methods in the last three rows
of Table~\ref{table:horserace} need little data structure bookkeeping
as space usage is $O(1)$.

When a new bucket is created, there are pros and cons to involve
``all'' vs ``few'' buckets to decide which keys move. At one end, when
involving all buckets (rendezvous, jump consistent), Lamping and Veach
observe that it is better to take few keys from each of them to
relieve a hot spot, but this requires many servers scanning and
sending data. At the other end (consistent, linear, round-), involving
few buckets may not relieve a hot spot soon, but it makes sense if the
data storage is distributed geographically among many data centers,
and the most efficient way to move data is to make a copy on physical
devices, moving them with a truck to the new data center. Note that
suitably increasing $s_0$ can combine the best of these two behaviors,
so the choice depends on the application domain.

\paragraph*{Our second contribution} 
We performed an experimental study of the above hashing
methods, since Table~\ref{table:horserace} does not give the full picture
from an algorithm engineering point of view. The code is publicly
available at \url{https://github.com/veluca93/round_hashing} to replicate the experiments.

Our first observation addresses how balanced are the buckets filled with the
hashing methods in Table~\ref{table:horserace}. By uniformly sampling
all the possible keys, their hash values can be used to estimate how
far the number of keys in buckets are from the ideal load factor
$\alpha$, reporting the least and the most populated buckets after the
experiments. We observed that jump consistent hashing is very close to
$\alpha$, ranging from $0.988 \, \alpha$ to $1.012 \, \alpha$; the
experimental study in~\cite{LampingV14} shows that it compares
favorably with consistent hashing (rendezvous hashing is not
directly compared). We can match this performance by setting $s_0 =128$ for
linear hashing and $s_0 = 64$ for round-hashing.

Our second observation relates to the real cost of instructions on a
commodity processor. Concretely to illustrate our points, we refer to
Intel processors~\cite{intel}. Here Euclidean division is not our
friend: integer division and modulo operations on 64-bit integers take
85--100 cycles, whereas addition takes 1 cycle (and can be easily
pipelined). Interestingly, this goes in the direction of the so-called
AC$^0$-RAM dictionaries (e.g. see Andersson et
al.~\cite{ACOdicitonary}) and Practical RAM (e.g. see Brodnik et
al.~\cite{BrodnikMM97} and Miltersen~\cite{Miltersen96}), where
integer division and multiplication are not permitted, among
others.\footnote{We thank Rasmus Pagh for pointing us the reference on
  AC$^0$-RAM dictionaries.} However, multiplication should be taken
with a grain of salt as, surprisingly, it takes 3--4 cycles (which
becomes 1 cycle when it can be pipelined). Also, the modulo operation
for powers of two or for small constants proportional to $s_0$, can be
replaced with a few shift and multiplication
operations~\cite{Gr.Mo:div:94} as available, for instance, in the
\texttt{gcc} compiler from version 2.6.  Our implementation of
round-hashing avoids general integer division and modulo operations
because they are almost two orders of magnitude slower than the other
operations: using them could nullify the advantage of the $O(1)$ time
complexity. We applied this tuning also to linear hashing whenever
possible.

As a result, to find the bucket number for a key, round-hashing is
almost an order of magnitude faster than jump consistent hashing, and
even much faster than the other hashing methods in
Table~\ref{table:horserace}.  This is crucial for the system
throughput: first, round-hashing can \emph{serve} tenfold or more requests;
second, when a new bucket number is added, it improves the performance of
\emph{rescanning} the keys to decide which ones move to the new bucket. We
refer the reader to Section~\ref{sec:load} for further details on our
experimental study.

\paragraph*{Our third contribution} 
We apply round-hashing to obtain a variant of dynamic hash tables,
called \emph{round-table}, that addresses the issues of a
high-throughput server with many lookup requests, relatively few
updates, and where some keys can be kept in a $\stash$ in main
memory. We follow the classical two-level external-memory
model~\cite{Aggarwal:1988:IOC} to evaluate the complexity. Let $n$ be
the number of keys currently stored in the table, and $B$ be the
maximum number of keys that fit inside one block transfer from main
memory to external memory, or vice versa. A $\stash$ of size $k$ keys
can be kept in main memory. We measure space occupancy using the \emph{space
  utilization} $1-\epsilon$, where $0 \leq \epsilon < 1$, defined as
the ratio of the number $n$ of keys divided by the number of
external-memory blocks times $B$, hence the number of blocks is
$\lceil \frac{n}{B(1-\epsilon)} \rceil$. In other words, $\epsilon$
represents the ``waste'' of space in external memory, so the lower
$\epsilon$, the better.

Round-table achieves the following bounds. Each lookup reads just
1~block from external memory in the worst case, taking $O(1)$ CPU time
and thus requiring only $O(1)$ words from main memory. Each update
(insertion or deletion) requires to access at most $4 s_0$ blocks in
external memory, in the worst case, taking
$O(s_0 (B+\log n/\log\log n))$ CPU time w.h.p.~(expected time is
$O(s_0 B)$) and using $O(B)$ memory cells.  The number of keys in the
$\stash$ is $k \approx n/\exp(B)$.  For example, setting
$s_0 = 2/\epsilon$ when $\epsilon >0$, we obtain
$k \approx n \frac{\exp[-{\frac{B}{4} \cdot
    \frac{\epsilon^2}{2-\epsilon}}]}{\sqrt{\pi B (2-\epsilon)}}$,
noting that the update cost becomes $O(\epsilon^{-1})$ in this case.
Thus $k$ is exponentially smaller than $\Omega(n/B)$ in main memory,
obtained by storing at least one word per external block (as B-trees
do). Experiments in Section~\ref{sec:round-table} confirm our
estimation.

Looking at the vast literature on hashing, apart from the previously
mentioned methods, Mirrokni et al.~\cite{45756} provide a version of
consistent hashing that keeps bucket load within a factor of
$1+\epsilon$, but it cannot guarantee at most one memory access.  The
expected optimal bound of $1+O(1/\sqrt{B})$ memory accesses in Jensen
and Pagh~\cite{JensenP08} does not require the stash, with no
guarantee of at most one memory access. As for the work on tables
with one external-memory access, some
results~\cite{mairson1983program,Fagin:1979:EHF:320083.320092} rely on
perfect hashing, but are either not dynamic or cannot reach
arbitrarily high utilization.  A recent cuckoo hashing based
approach~\cite{DBLP:journals/corr/abs-1709-04711}, combined with
in-memory Bloom filters to ensure that lookups access the correct
position, is not simple to dynamize. A general
scheme~\cite{Ramakrishna1989} relies on perfect hashing to store the
$\stash$ on external memory, thus having higher worst-case cost for
insertions.  The result in~\cite{Cesarini1993} achieves single-access
lookups, but at the cost of $O\left(\frac n {B(1-\epsilon)}\right)$
internal memory.  A solution based on predecessor search needs
$O\left(\frac n B\right)$ internal memory, as discussed
in~\cite{pagh2003basic}.

\section{Round-Hashing and Round-Mapping}%
\label{sec:round-mapping}

Round-hashing maps the hash value of the input key into a position $u$
along the unitary circumference $C$, and invokes round-mapping on $u$
to find the corresponding bucket number (note that $u$ can be seen as
a fraction of the unit). To this end, given the integer slack
parameter $s_0 > 0$, round-mapping maintains a partition of $C$ into
arcs of length proportional to either $1/s$ or $1/(s+1)$ for some
integer $s$ ($s_0 \leq s \leq 2s_0-1$): if there are $m$ arcs in $C$,
they are consecutively numbered from $0$ to $m-1$ in clockwise order
and in one-to-one correspondence with a permutation of
$\{0, 1, \dots, m-1\}$, called the \emph{bucket numbers}. Also we
refer to arcs as \emph{short} when their length is proportional to
$1/(s+1)$, and \emph{long} when their length is proportional to
$1/s$. Round-mapping supports the following operations, which are
better visualized using $C$ and its arcs in clockwise order.

\begin{itemize}
\item $\init()$: divide the circumference $C$ into $s=s_0$ long arcs, and set $m=s_0$.
\item $\numBuckets()$: return the current number $m$ of arcs (and, thus, of
      buckets).
\item $\findBucket(u)$ for $0 \leq u < 1$: return the bucket number assigned to
     the arc hit by the clockwise fraction $u$ of the circumference $C$,
     where $u=0$ hits the arc~$0$.
\item $\newBucket()$: if all arcs are short, make them long; also, if
     $s = 2*s_0-1$ then set $s := s_0$, else $s := s+1$. Starting from
     arc~$0$ and proceeding clockwise in $C$, find the sequence of the
     $s$ closest long arcs, and shrink each of them so that it becomes
     short. 
     In this way, a new short arc is allocated at the end of the
     sequence, and is associated with a new bucket number $m$ (i.e.
     $\numBuckets() = m+1$).  Return the bucket numbers for the former
     arcs.
   \item $\freeBucket()$: inverse operation of $\newBucket()$. If all
     arcs are long, make them all short; also, if $s = s_0$ then set
     $s := 2*s_0-1$, else $s := s-1$. Take the $s+1$ closest short
     arcs in clockwise order, discard the last one, and change each of
     the first $s$ ones to be long, releasing the largest bucket
     number $m-1$ (associated with the discarded arc, hence
     $\numBuckets() = m-1$).  Return the bucket numbers of the
     original $s+1$ short arcs.
\end{itemize}

Wlog we focus on $\newBucket()$, as $\freeBucket()$ is simply
unrolling the last $\newBucket()$ operation. They both require
$O(s_0)$ time in the worst case, which is $O(1)$ when $s_0=O(1)$.

We say that a \emph{round} starts when $s=s_0$. Let the rounds be numbered as
$q = 0, 1, 2, \ldots$, and let the length $\len(q) = s_0 \, 2^q$ of a
round~$q$ represent how many bucket numbers have been allocated by the
last step $s = 2 s_0-1$.  For example, choosing $s_0=3$, the first
rounds are shown in Figure~\ref{fig:example-round-mapping}.  At step
$s=4$ of round $q=4$, each call to $\newBucket()$ takes $s$ consecutive
bucket numbers and inserts a new bucket number: after 0 1 2 24 it
inserts 32, after 12 16 20 25 it inserts 33, after 6 8 10 26 it
inserts 34, and so on. Note that 32, 33, 34, etc., are \emph{native}
of round~$q=4$. Figure~\ref{fig:example-round-mapping} shows of
which round the bucket numbers are native, using framed chunks. After
the last step $s=2s_0-1$, each round contains twice the bucket numbers
than after the last step of the previous round. Also, the concatenation of
every other chunk of $s_0$ non-native bucket numbers, produces exactly the
outcome of the previous round. We will exploit this regular pattern.

As it can be noted, a mapping between the arcs and a permutation is
maintained: for example, in round $q=4$, arc $j$ for $j=0,1,2$ corresponds to
bucket number $b(j)=j$; arc $j=4$ has $b(j)=24$, arc $j=5$ has $b(j)=32$,~\dots,
and arc $j=47$ has $b(j)=47$. Note that we do not maintain this mapping
explicitly; still, $\findBucket(u)$ is able to identify arc~$j$ and its bucket
number $b(j)$ in constant time.

\begin{algorithm}[t]
\small
\DontPrintSemicolon%
\SetKwProg{myproc}{Function}{}{}
  \myproc{$\findBucket(u)$}{
    $j \gets$ arc hit by $u$\;
    \lIf{$j < s_0$}{\Return{$j$}}
    \lIf{$j > p$}{$j' \gets j - \frac{p+1}{s+1}$, $s'=s$}
    \lElse{$j'\gets j$, $s'=s+1$}
    $x \gets (j' \modulo s') \modulo s_0$\;
    $q' \gets q + \left\lfloor \frac{s'-1}{s_0} \right\rfloor$\;
    $i = \left(1+ \left\lfloor \frac{s'-1}{s_0} \right\rfloor\right)
    \cdot \left\lfloor \frac{j'}{s'}  \right\rfloor 
    + \left\lfloor \frac{j' \modulo s'}{s_0} \right\rfloor$\;
    \Return{$\pos(i,x,q')$}
  }

\BlankLine%

\SetKwProg{myproc}{Function}{}{}
  \myproc{$\pos(i,x,q)$}{
    $e \gets$ position of the least significant bit~1 in $i$\;
    \Return{$\lfloor \frac{(s_0+x) 2^q + i}{2^{e+1}}\rfloor$}
}

\caption{Mapping from arcs to buckets}%
\label{alg:find-bucket}
\end{algorithm}

\subsection{Implementation of $\findBucket(u)$}

We exploit the invariant property that short arcs are numbered from $0$ to $p$,
and thus $p+1$ is a multiple of $s+1$, where $p$ is maintained as the last added
short arc. We also use $\pow(a)$, where $a>0$, to denote the largest integer
exponent $e \geq 0$ such that $2^e$ divides $a$ (a.k.a.~2-adic order).
Equivalently, $\pow(a)$ is the position of the least significant bit~1 in the
binary representation of the unsigned integer $a > 0$.

First, consider the ideal situation: after the last step $s=2s_0-1$ of
round $q$, we have $\len(q)$ buckets, numbered
consecutively from $0$ to $\len(q)-1$. We also have $\len(q)$ arcs on the
circle, numbered consecutively from $0$ to $\len(q)-1$. As arc $j$ is mapped to
bucket number $b(j)$ using our scheme, we give a closed formula for $b(j)$ that
can be computed in $O(1)$ time in the word RAM model, where divisions
and modulo operations involve just powers of two or constants in
the range $[s_0 \ldots 2 s_0]$.

Let $j = s_0 \, i + x$ where $x \in \{0,1,\ldots, s_0-1\}$. 
If $i=0$, then $b(j) = b(x) = x$. Thus $b(j) = j$ for
$0 \leq j < s_0$.  Hence, let assume $i>0$ in the rest of the section,
and thus we need to compute $b(j)$ for $j \geq s_0$.  

We say that the bucket number in position $j$ belongs to \emph{chunk} $i$ (hence,
a chunk is of length $s_0$). For odd values of $i$, the bucket number
is native for round $q$. For even values of $i$, the bucket number is
native for round $q-\pow(i)$, as it can be checked in
Figure~\ref{fig:example-round-mapping}: for example, in round $q$
after the last step, bucket number~$9$ is in position
$j=37= 3 \cdot 12 + 1$, so $i=12$ and $9$ is native for round
$q-\pow(i) = 4-2 =2$. In general, as $\pow(i) = 0$ when $i$ is odd, we
can always say that the bucket number is native for round $q-\pow(i)$
for $i>0$. Another useful observation is that the smallest native
number in round $q$ is $\len(q-1)$ by construction (e.g.~24 in round
$q=4$).

In the ideal situation, we find the native round for the bucket number
at position $j$: as its chunk is preserved in the native round, we can
use its offset $x$ inside the chunk to recover the value of that bucket
number. In the native round $q$, each chunk $i$ starts with bucket
number $\len(q-1)$ as previously observed, increased by one for each
such chunk, thus the first bucket number in chunk $i$ is
$\len(q-1) + \lfloor i/2 \rfloor$. Also, any two adjacent numbers in
the chunk, differ by $2^{q-1}$ by construction.  Summing up, there are
two cases for the bucket number for $j$:
\begin{itemize}
\item $i$ odd and thus native for round $q$: the bucket number is
  $\left\lfloor \frac{(s_0+x) 2^q + i}{2} \right\rfloor$
\item $i$ even and thus native for round $q-\pow(i)$: the bucket number
  is $\left\lfloor \frac{(s_0+x) 2^q + i}{2^{\pow(i)+1}} \right\rfloor$
\end{itemize}
As $\pow(a) = 0$ when $a$ is odd, we can compactly write these
positions in the ideal situation as
\[
\pos(i,x,q) = \left\lfloor \frac{(s_0+x) 2^q + i}{2^{\pow(i)+1}}
\right\rfloor
\]

Second, consider the general situation, with an intermediate step
$s_0 \leq s \leq 2 s_0 -1$ in round $q$. Recall that we know the
position $p$ of the last created arc. This gives the following
picture. The first $p+1$ short arcs in clockwise order can be seen as
$\frac{p+1}{s+1}$ consecutive groups, each of $s+1$ arcs, and the
remaining arcs are long and form groups of $s$ arcs each. Let us set
$s' = s+1$ in the former groups, and $s'= s$ in the latter groups.  In
the following, we equally say that each group contains $s'$ arcs or
that each group contains $s'$ bucket numbers. In general, we say $s'$
entries (arcs or bucket numbers) when it is clear from the context.
 
A common feature is that the first $s_0$ entries of each group are
inherited from the previous round, and the last $s'-s_0$ entries in
each group are those added in the current round: each new entry is
\emph{appended} at the end of each group, so the entry in position $p$
is the last in its group.

Now, given a position~$j$, we want to compute $b(j)$, the
corresponding bucket number. The idea is to reduce this computation to
the ideal situation analyzed before. 

If $j > p$, we conceptually remove one entry for each group such that
$s' = s+1$. This is equivalent to set $j := j - \frac{p+1}{s+1}$ and,
consequently, $p := p - \frac{p+1}{s+1}$. Now, we conceptually have
all the groups of the same size $s'$, which are sequentially numbered
starting form $0$.

Let $i' = \left\lfloor j/s' \right\rfloor$ be the number of the group
that contains the entry corresponding to $j$. We now decide whether
$j$ is one of the first $s_0$ entries of its group or not. We have two
cases, according to the value of $r = j \modulo s'$.

If $r < s_0$, the wanted entry is one of the first $s_0$ entries of its
group. If we concatenate those entries over all groups, we obtain the
ideal situation of the previous round $q-1$. There, the wanted entry
occupies position $j' = s_0 i' + r$. Hence, $b(j) = \pos(i', r, q-1)$
in the ideal situation.

If $r \geq s_0$, the wanted entry is one of the last $s'$ entries of its
group. Analogously, if we concatenate those entries over all groups, the
position of the wanted entry becomes $j'' = (s'-s_0) i' + r - s_0$,
where $x = r - s_0$ is the internal offset. However, we cannot solve
this directly. We use instead the observation that the futures entries
that will contribute to get the ideal situation for round $q$, will be
\emph{appended} at the end of each group. In this ideal situation, the
wanted entry correspond to arc $2i'+1$ and is at position
$j' = s_0 (2i'+1) + r - s_0$ for round~$q$. Thus,
$b(j) = \pos(2i'+1, r-s_0, q)$ in the ideal situation.

We can summarize the entire computation of $b(j)$ in an equivalent
formula computed by Algorithm~\ref{alg:find-bucket}
that can be computed in $O(1)$ time.

\begin{lemma}%
  \label{lemma:findBucket}
  $\findBucket()$ can be implemented in $O(1)$ time using bitwise operations.
\end{lemma}

Interestingly, $\findBucket()$ is much faster then other approaches known in the
literature for consistent hashing, as we will see in Section~\ref{sec:load}. 

\begin{theorem}%
  \label{the:round-mapping}
  Round-mapping with integer parameter $s_0 >1$ can be implemented
  using $O(1)$ words, so that $\init()$, $\numBuckets()$ and
  $\findBucket()$ take $O(1)$ time, and $\newBucket()$ and
  $\freeBucket()$ take $O(s_0)$ time.
\end{theorem}

\begin{figure*}[t]
  \centering
  \includegraphics[width=\textwidth]{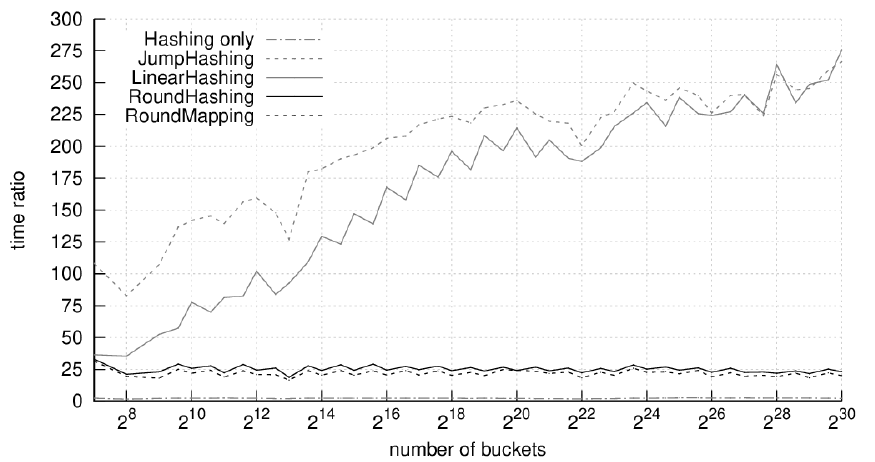}
  \caption{Time needed to compute a single hash as the number of buckets varies.}\label{fig:hashtime}
\end{figure*}

\section{Distributed Servers}%
\label{sec:load}
We experimentally evaluated round-hashing and our C~implementation of
Algortihm~\ref{alg:find-bucket}, on a commodity hardware based on Intel
Xeon E3-1545M v5 CPU and 32Gb RAM, running Linux 4.14.34, and using
\texttt{gcc}~7.3.1 compiler. We give some implementation details on
the experimented algorithms, observing that we decided not
to run consistent hashing~\cite{Karger:1997:CHR:258533.258660} and
rendezvous hashing~\cite{ThalerR98} as they are outperformed by jump
consistent hashing as discussed in detail
in~\cite{LampingV14}. Specifically, we ran the following code.
\begin{itemize}
\item Jump consistent hashing~\cite{LampingV14}: we employed the
  implementation provided by the authors' optimized code.
\item Linear hashing~\cite{larson1980linear,litwin1980linear}: the
  pseudocode is provided but not the code, which we wrote in~C. As for
  the $O(\log m)$ hash functions, we followed the approach suggested
  in~\cite{larson1980linear}: we employed the fast and high-quality
  pseudo-random number generator in~\cite{xoroshiro128+} using the key
  to hash as a seed and the $j$th output as the outcome of the $j$th
  hash function. This takes constant time per hash function. Moreover,
  we replaced all modulo operations with the equivalent faster
  operations, as we did for round-mapping.
\item Round-hashing (this paper): we employed the first output from
  the pseudo-random number generator in~\cite{xoroshiro128+} as hash
  value.  We chose the size of our hash range to be a power of two,
  so that mapping a hash value to an arc number can be done without
  divisions: we computed the product between the number of buckets and
  the hash value, divided by the maximum possible hash value. Note
  that some care is required to compute the product correctly as it
  may overflow.
\end{itemize}

It is worth noting that replacing the expensive division was very
effective in our measurements. In particular, we replaced the division
by $s'$ in Algorithm~\ref{alg:find-bucket} with the pre-computed
equivalent combination of multiplication and shift: as
$s_0 \le s' \leq 2s_0$, this can be done at initialization time with a
constant amount of work. This reduced the time per round-hashing call
from $14.02$ns to $8.71$ns, a $60\%$ decrease, which is an interesting
lesson that we learned.

Figure~\ref{fig:hashtime} shows the running times for the above
implementations, when computing ten million hash values, as the number
of buckets varies on the x-axis. On the y-axis, the running times are
reported for jump consistent hashing, linear hashing, and three
versions of our round-hashing: the full round-hashing cost (i.e. given
a key, return its bucket number); the cost of round-mapping alone
(i.e. given a position $u$ in the circumference, return its bucket
number); and the cost of computing just the hash value
using~\cite{xoroshiro128+}. As it can be seen, as the overhead of the
latter is negligible, the costs of round-hashing and round-mapping are
very close and \emph{constant} along the x-axis, outperforming the
non-constant costs of jump consistent hashing and linear hashing, which
behave similarly when the number of buckets is large. Note that
round-hashing has at least an order of magnitude improvement at
around $2^{16}$ buckets and on, which indicates that it scales well.

All the running times in Figure~\ref{fig:hashtime} were normalized by
the time needed to compute the sum of all the values. Looking at the
absolute figures, the running time for the sum is about $0.4$ns per
element, and that of round mapping is $8$--$10$ns per element (and the
pseudo-random number generator in~\cite{xoroshiro128+} takes twice the
cost of the sum).

Speed is not the whole story as it is important also how the hash
values in the range are distributed in the buckets. To this end, we
show in Table~\ref{tab:distrstats} the results using $64$-bit hash
values: as it was infeasible to compute the bucket for every possible
hash value, we chose $10^9$ values at regular intervals in the hash
range of $2^{64}$ values, and computed the bucket size distribution for
them.

The columns in the table report the parameters for $10^4$ buckets,
where the actual bucket sizes are obtained by multiplying parameters
in \{min,max,1\%,99\%\} by the load factor $\alpha= 10^9/10^4$.
Specifically, $s_0$ useful for linear hashing and round-hashing, the
standard error $\frac{\sigma}{\mu}$ where $\sigma$ is the variance and
$\mu$ is the average of the bucket sizes, the minimum and maximum
bucket size, the 1\% and 99\% percentiles of the size, and the ratio
between the latter two. This ratio is the most important parameter in
the table as it shows how well-balanced are buckets. It can be easily
seen that both round-hashing and linear-hashing can match almost
perfectly, with round-hashing having a slightly better distribution.
Based on this table, we can see that round-hashing and linear-hashing
have distribution properties that are similar to jump consistent
hashing, as long as we choose suitable values: $s_0 = 64$ for
round-hashing and $s_0 = 128$ for linear hashing.
Figure~\ref{fig:hashtime} has been plotted using these values of
$s_0$.

\begin{table*}[t]
  \centering
\begin{tabular}{|rr|c|cc|cc|c|}
  \hline
            & $s_0$ & $\frac{\sigma}{\mu}$ &   min   &   max   &   1\%    &   99\%   & percentile ratio \\
  \hline
  \multicolumn{2}{|c|}{jump consistent h.} &       $0.316$        & $0.988$ & $1.012$ & $0.993$ & $1.007$ &   $1.014$  \\
  \hline
  round-hashing &  $1$  &      $29.325$        & $0.610$ & $1.221$ & $0.610$ & $1.221$ &   $2.001$  \\
            &  $2$  &      $20.272$        & $0.814$ & $1.221$ & $0.814$ & $1.221$ &   $1.500$  \\
            &  $4$  &       $7.192$        & $0.977$ & $1.221$ & $0.977$ & $1.221$ &   $1.250$  \\
            &  $8$  &       $4.465$        & $0.976$ & $1.085$ & $0.976$ & $1.085$ &   $1.112$  \\
            &  $16$ &       $2.560$        & $0.976$ & $1.028$ & $0.976$ & $1.028$ &   $1.053$  \\
            &  $32$ &       $0.613$        & $0.976$ & $1.002$ & $0.976$ & $1.002$ &   $1.027$  \\
            &  $64$ &       $0.421$        & $0.989$ & $1.002$ & $0.989$ & $1.002$ &   $1.013$  \\
            & $128$ &       $0.277$        & $0.995$ & $1.002$ & $0.995$ & $1.002$ &   $1.007$  \\
  \hline
  linear hashing &  $1$  &      $29.329$        & $0.602$ & $1.232$ & $0.605$ & $1.228$ &   $2.030$  \\
            &  $2$  &      $20.274$        & $0.803$ & $1.234$ & $0.808$ & $1.228$ &   $1.520$  \\
            &  $4$  &       $7.203$        & $0.964$ & $1.232$ & $0.969$ & $1.225$ &   $1.264$  \\
            &  $8$  &       $4.476$        & $0.965$ & $1.095$ & $0.970$ & $1.090$ &   $1.124$  \\
            &  $16$ &       $2.583$        & $0.965$ & $1.041$ & $0.970$ & $1.034$ &   $1.066$  \\
            &  $32$ &       $0.685$        & $0.968$ & $1.014$ & $0.973$ & $1.009$ &   $1.037$  \\
            &  $64$ &       $0.527$        & $0.980$ & $1.014$ & $0.984$ & $1.009$ &   $1.025$  \\
            & $128$ &       $0.417$        & $0.985$ & $1.014$ & $0.990$ & $1.009$ &   $1.019$  \\
  \hline
  \multicolumn{8}{c}{} \\
\end{tabular}
\caption{Statistics on how much hash space is assigned to a given bucket,
  with a total of $10000$ buckets. Note that the actual bucket sizes are
  obtained by multiplying the numbers in columns min, max, 1\%, 99\&
  by the load factor $\alpha$. Extremal values and percentiles are a ratio
  from the ideal value.}\label{tab:distrstats}
\end{table*}

\begin{table*}
\hspace*{-18pt}
\begin{tabular}{|r|rr|rr|rr|rr|rr|rr|}
  \hline
  \multicolumn{1}{|c|}{\multirow{3}{*}{$s_0$}}  & \multicolumn{12}{c|}{$\epsilon$} \\
  \cline{2-13}
  & \multicolumn{2}{c|}{\small 0} & \multicolumn{2}{c|}{\small 0.001} & \multicolumn{2}{c|}{\small 0.01} & \multicolumn{2}{c|}{\small 0.03} & \multicolumn{2}{c|}{\small 0.05} & \multicolumn{2}{c|}{\small 0.1} \\
  & \multicolumn{1}{c}{real} & \multicolumn{1}{c|}{est.} & \multicolumn{1}{c}{real} & \multicolumn{1}{c|}{est.} & \multicolumn{1}{c}{real} & \multicolumn{1}{c|}{est.} & \multicolumn{1}{c}{real} & \multicolumn{1}{c|}{est.} & \multicolumn{1}{c}{real} & \multicolumn{1}{c|}{est.} & \multicolumn{1}{c}{real} & \multicolumn{1}{c|}{est.} \\
  \hline
  1 & \small 17.2\% & \small 18.4\% & \small 17.1\% & \small 18.3\% & \small 16.7\% & \small 17.4\% & \small 15.9\% & \small 15.7\% & \small 15.1\% & \small 14.5\% & \small 13\% & \small 12\% \\
  4 & \small 5.6\% & \small 6.8\% & \small 5.5\% & \small 6.7\% & \small 5.1\% & \small 5.9\% & \small 4.2\% & \small 4.4\% & \small 3.4\% & \small 3.4\% & \small 1.7\% & \small 1.7\% \\
  16 & \small 1.8\% & \small 2.8\% & \small 1.7\% & \small 2.7\% & \small 1.3\% & \small 1.9\% & \small 0.7\% & \small 0.7\% & \small 0.3\% & \small 0.1\% & \bf \small 0.01\% & \bf \small 0.1\% \\
  32 & \small 1.4\% & \small 2\% & \small 1.3\% & \small 1.9\% & \small 0.9\% & \small 1.2\% & \small 0.4\% & \small 0.3\% & \bf \small 0.1\% & \bf \small 0.5\% & \bf \small 0.003\% & \bf \small 0.009\% \\
  64 & \small 1.3\% & \small 1.6\% & \small 1.2\% & \small 1.5\% & \small 0.8\% & \small 0.9\% & \bf \small 0.3\% & \bf \small 0.6\% & \bf \small 0.1\% & \bf \small 0.2\% & \bf \small 0.003\% & \bf \small 0.002\% \\
  256 & \small 1.3\% & \small 1.3\% & \small 1.2\% & \small 1.3\% & \bf \small 0.8\% & \bf \small 1\% & \bf \small 0.3\% & \bf \small 0.3\% & \bf \small 0.08\% & \bf \small 0.09\% & \bf \small 0.003\% & \bf \small 0.0005\% \\
\hline
  ideal & \multicolumn{1}{c}{-} & \small 1.2\% & \multicolumn{1}{c}{-} & \small 1.2\% & \multicolumn{1}{c}{-} & \small 0.8\% & \multicolumn{1}{c}{-} & \small 0.3\% & \multicolumn{1}{c}{-} & \small 0.07\% & \multicolumn{1}{c}{-} & \small 0.0003\% \\
  \hline
  \multicolumn{13}{c}{} \\
\end{tabular}
\caption{Percentage of elements on the $\stash$ as $s_0$ and $\epsilon$
change, with $B=1024$.}\label{tab:B1024}
\end{table*}
  
\section{External-Memory Tables}%
\label{sec:round-table}

Given a universe $U$ of keys, and a random hashing function
$h : U \rightarrow I$, where $I = \{ 0, 1, \ldots, |I|-1 \}$, we build
a hash table that keeps a $\stash$ of keys in main memory. Armed with
the round-hashing, we obtain a hash table called \emph{round-table}
that uses $O(k+1)$ words in main memory, where $k$ denotes the number of stash
keys. We consider the $\stash$ to be a set of $k$ keys, where notation
$\stash[b]$ indicates the set $\{ x \in \stash : \findBucket(h(x)/|I|) = b \}$
(e.g. a hash table in main memory with maximum size $O(B+\log n/\log \log n)$
w.h.p. via a classical load balancing argument). To check if $x \in \stash$, we
check if $x \in \stash[b]$ where $b = \findBucket(x)$.
Also, for a user given parameter $\epsilon$, the guaranteed space utilization
in external memory is $1-\epsilon$.

The lookup algorithm is straightforward while
the insertion algorithm is a bit more complex (see
the appendix for the pseudocode). After checking that the key is not in the
table, it proceeds with the insertion. For this, we need to maintain
the claimed space utilization of $(1-\epsilon)$. That is, if
$\lceil \frac{n}{B (1-\epsilon)}\rceil > \numBuckets()$, we need one
more block. We invoke $\newBucket()$, and receive a list of
$z < 2s_0$ block numbers. We have to distribute the keys stored in
these $z$ blocks over $z+1$ blocks, where the extra block has
number $\numBuckets()$ as it is the latest allocated block number by
round-mapping.  In the distribution, the keys from the $\stash$ are
also involved, as described  below in the function $\distribute$. After that,
$\findBucket()$ finds the external-memory block $\block()$ that should
contain the key: if it is full, the key is added to the $\stash$.

Function $\distribute(b_0, b_1, \ldots, b_{z-1})$ takes these
$z$ block numbers from $\newBucket()$, knowing that
$b_z = \numBuckets()$ is the new allocated block number, and thus
allocates $\block(b_z)$.  Then it loads $\block(b_{z-1})$ and moves to
$\block(b_z)$ all keys $x \in \block(b_{z-1})$ such that
$\findBucket(x) = b_z$. Also, for each $x \in \stash[b_{z-1}]$ such
that $\findBucket(x) = b_z$, it moves $x$ to $\block(b_z)$, if there is
room, or to $\stash[b_z]$ otherwise.  Next, we repeat this task for
$b_{z-2}$ and $b_{z-1}$ while also taking care of moving keys from
$\stash[b_{z-1}]$ to $\block(b_{z-1})$ if there is room, and so on. In
this way, the cost of $\distribute$ is $2z+1$ block transfers, using
$O(B)$ space in main memory, taking $O(s_0 (B +\log n/\log \log n))$
CPU time w.h.p., and $O(s_0 B)$ expected time.

The deletion algorithm is similar to the insertion one (see the
appendix for the pseudocode), and its performance can be bound in the
same way as above. We check the condition
$\lceil \frac{n}{B (1-\epsilon)}\rceil < \numBuckets() - 1$ for
$n > 0$ to run $\freeBucket()$ using a slightly different
$\distribute$ that proceeds in reverse. Note that the rhs of the
condition is $\numBuckets() - 1$ to avoid $\newBucket()$ being called
too soon.

In Appendix~\ref{sub:stashround}, we show that as long as we choose $s_0 >
\frac2\epsilon$ we have that the stash size of a hash table implemented with
round-hashing is similar to the behaviour we would get with an uniform hash
function (that would require rehashing). Thus, we recommend choosing
$s_0\epsilon > 2$, as confirmed by the experiments below.
Moreover, in Appendix~\ref{sec:ext-stash} we show how to keep a copy of the
stash in external memory, without increasing space usage but increasing the
number of block operations per update to $O(1+\epsilon s_0)$.

To evaluate our approach, we consider the worst-case stash size (over
the number of keys) across multiple values of $n$ (going from
$2^{10}B$ to $2^{13}B$) for $B=512, 1024, 2048$ as $\epsilon$ and
$s_0$ vary. The results are reported in
Tables~\ref{tab:B1024},~\ref{tab:B512}~and~\ref{tab:B2048} (the last two can be
found in the appendix), where the left side of every column reports the ratio
predicted by the analysis of Appendix~\ref{sub:stashround} and the right side
shows the effective maximum ratio reported during the experiment. As our
analysis is substantially different when $\epsilon s_{0} > 1$, we reported
those values in bold to highlight them. Finally, the last row reports the best
values one can hope to achieve for that value of $\epsilon$, that is, the
values that our analysis predicts for a uniform hash function.

Looking at these results, we can make some observations. First, the
values predicted by the analysis match the results fairly well,
especially when $s_0\epsilon \gg 1$ or $s_0\epsilon \ll 1$. In
particular, it almost never happens that the analysis is wrong by more
than a factor of $3$. Second, when $s_0$ is small, stash size is
fairly high, even for low space utilization. This is to be expected,
as in this case different buckets may have very different assignment
probabilities. Third, as $s_0$ grows, stash size quickly approaches
the one that we would expect from the ideal case. Nonetheless, the
improvement is fairly small when $s_0$ goes over $32$, even at low
utilization.  We thus recommend $s_0$ to be chosen near $32$ for
practical usage.

\begin{figure*}[tb]
  \centering
  \begin{subfigure}{0.4\textwidth}
    \includegraphics[width=\textwidth]{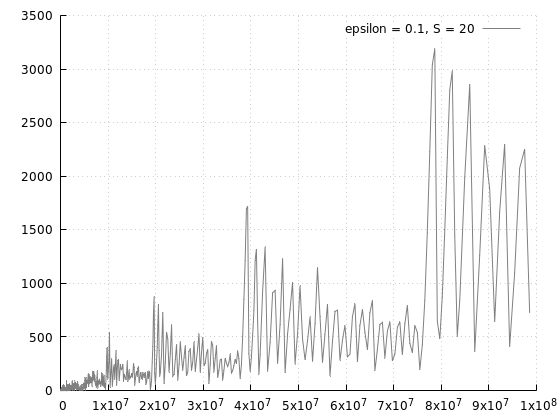}
    \caption{$\epsilon=0.1$}
  \end{subfigure}
  \begin{subfigure}{0.4\textwidth}
    \includegraphics[width=\textwidth]{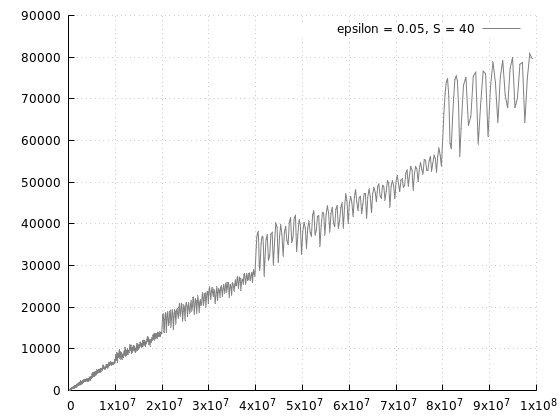}
    \caption{$\epsilon=0.05$}
  \end{subfigure}
  \begin{subfigure}{0.4\textwidth}
    \includegraphics[width=\textwidth]{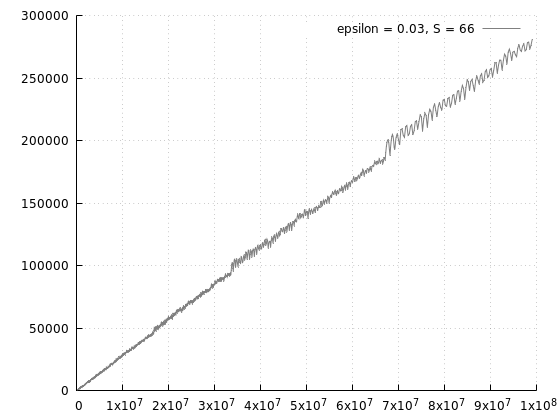}
    \caption{$\epsilon=0.03$}
  \end{subfigure}
  \begin{subfigure}{0.4\textwidth}
    \includegraphics[width=\textwidth]{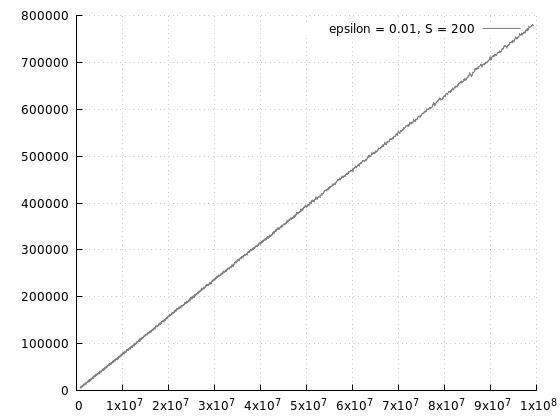}
    \caption{$\epsilon=0.01$}
  \end{subfigure}
  \caption{Stash size (on the y-axis) as $n$ grows (on the x-axis) for $s_0=\frac{\epsilon}2$ and different values of $\epsilon$.}\label{fig:stashsize}
\end{figure*}

We also considered how stash size varies over time, as more elements are
inserted. To study that, we fixed $s_0 = \frac{2}{\epsilon}$, as recommended in
the analysis section, and plotted the size of the $\stash$ against the number of
elements in the table. The plots can be found in Figure~\ref{fig:stashsize}.
These plots clearly show the ``cyclic'' behavior of round-table: when a new
round begins, the distribution of keys in buckets is further away from being
uniform and, as a result, the stash size increases. As more steps of the
round are completed, the spikes in stash size get progressively smaller as
round-table balances keys in a better way, until a new round starts again and
the table reverts to its previous behavior.

\bibliographystyle{plain}
\bibliography{references} 

\appendix

\section{Analysis of stash size with a uniform hash function}%
\label{sub:stashunif}
In the following, we consider the arcs along the circle handled by
round-mapping as buckets.  To give bounds on the expected stash size for a hash
function that does not map a value into a bucket equiprobably~\cite{GonnetL88},
we will first study the expected number of elements that overflow from a bucket
of size $B$ that is expected to reach a load factor of $1-\delta$, i.e. when
any of the $n$ values are mapped to that bucket with probability $O\left(\frac
{B(1-\delta)}{n}\right)$. We will consider the cases in which $\delta = 0$,
$\delta < 0$ and $0 < \delta < 1$ separately.

The number of values that are assigned to the given bucket is a random variable
with binomial distribution $B\left(n, \frac{B(1-\delta)}{n}\right)$, which, as
$n$ grows, can be approximated with a normal variable with distribution
$N(B(1-\delta), B(1-\delta))$. The number of values that overflow from that
bucket is given by 

\begin{enumerate}
    \item $0$ if $x < B$ values end up in that bucket.
    \item $x-B$ if $x \ge B$ values end up in that bucket.
\end{enumerate}

From this follows that the expected number of overflown values from the bucket
is given by
\[
\frac{1}{\sqrt{2\pi B(1-\delta)}}
\int_{B}^{\infty}(x-B)e^{\frac{{(x-B(1-\delta))}^2}{2B(1-\delta)}}dx
\]

With some calculations, we find out that the value of this integral is given by
\[
  \frac{\sqrt{B(1-\delta)}e^{-\frac{B}{2}\frac{\delta^2}{1-\delta}}}{\sqrt{2\pi}}
  - \frac{\delta B}{2}
  \erfc\left(\sqrt{B}\frac{\delta}{\sqrt{2(1-\delta)}}\right)
\]
where $\erfc$ is the complementary error function.

\subsection{Case $\delta = 0$}
In this case, we easily get that the expected number of values that overflow
from a single bucket is $\sqrt{\frac{B}{2\pi}}$.

We can use this result to bound the stash size in the case $\epsilon = 0$ with
a uniform hash function: indeed, we get that the expected size of the $\stash$
grows as $ \frac{n}{\sqrt{2\pi B}}$.

\subsection{Case $\delta > 0$}
As $0 < \erfc(x) < 2$, we remark that the first term of the above expression
is an upper bound. This, as before, proves that the size of the $\stash$ with an
uniform hash function grows at most as
\[ n \frac{e^{-\frac B2 \frac{\epsilon^2}{1-\epsilon}}}{\sqrt{2\pi B(1-\epsilon)}} \]
which decreases exponentially in $B$. Under the assumption that
$\epsilon\sqrt{B}$ is big enough, we can replace $\erfc$ in the above expression
with its Taylor series to get a bound of
\[ n \frac{\sqrt{1-\epsilon}}{\epsilon^2 B\sqrt{2\pi B}}
e^{-\frac{B}{2}\frac{\epsilon^2}{1-\epsilon}} \] 
that improves the previous one by a factor of $\frac 1 B$. Of course, this
second bound only holds when $B$ and $\epsilon$ are large enough.

\subsection{Case $\delta < 0$}
In this case, we can replace $\erfc$ with its upper bound $2$ and ignore the
first term, as it is quite smaller than the second one. By multiplying by the
number of buckets as in the previous cases, we get the bound of  
$n \frac{-\delta}{1-\delta}$
for a hash table that tries to fit $n$ values in buckets that only have space
for $\frac{n}{1-\delta}$, as one would expect. This corresponds to an expected
stash size per bucket of $-\delta B$.

\section{Analysis of stash size with round-hashing}%
\label{sub:stashround}

Our version of consistent hashing guarantees that, at any time, the expected
load on the most-loaded bucket will be, at most, $1+\frac{1}{s} \leq
1+\frac{1}{s_0}$ times the expected
load on the least-loaded one. We can give an upper bound for the load
factor of those buckets as $(1-\epsilon)(1+\frac{1}{s_0}) \le 1-\epsilon + \frac{1}{s_0}$.
If $s_0 > \frac{2}{\epsilon}$, our hash table implemented with round-mapping this version of
consistent hashing will behave at least as well as a hash table implemented
with uniform hashing and a load factor of $1-\frac{\epsilon}2$, in terms of
stash size.

We will now consider what happens when $s_0 \le \frac{1}{\epsilon}$. For
simplicity, we will first study the behavior of the $\stash$ in the first step
$s=s_0$  of every round.  Let $c$ be the number of buckets that was present the
last time the buckets were all equally sized, and let $c+q$ be the current
number of buckets. We know that there are $c-s_0q$ buckets with size
proportional to $\frac{1}{c}$ and $(s_0+1)q$ buckets with size proportional to
$\frac{s_0}{c(1+s_0)} = \frac{1}{c+\frac{c}{s_0}}$. As the second kind of
buckets is less loaded than they should be with a uniform hash function, we
will ignore them as they will not contribute to the $\stash$ more than the
uniform case.

We expect each of the bigger buckets to be assigned $\frac{n}{c}$ keys, while
only having space for $B$. Thus, the $c-s_0q$ bigger buckets will behave as
buckets that have a load factor of $1-\delta$ with $\delta$ given by 
\[ -\delta = \frac{\frac{n}{c} - B}{B} = \left(1+\frac qc\right)(1-\epsilon)-1
\approx \frac qc - \epsilon\]
where we used the fact that $c+q = \frac{n}{B(1-\epsilon)}$. The analysis in
Subsection~\ref{sub:stashunif} gives us an expected stash size of
\[(c-s_0q)\left(\frac qc - \epsilon\right)B = n(1-\epsilon)\frac{(1-s_0\frac
qc)(\frac qc - \epsilon)}{1+\frac qc } \]

Since $c > s_0q$, we can use standard tools from analysis to find that the maximum
value of this function is realized when
\[ \frac qc = \sqrt{\left(1+\frac{1}{s_0}\right)(1+\epsilon)}-1 \]
with a value of
\[ n(1-\epsilon)\left[1+\epsilon s_0 - 2s_0\left(\sqrt{\left(1+\frac
1s_0\right)(1+\epsilon)}-1\right)\right] \]

Note that this expression is decreasing in $s_0$: this proves that the
worst-case behavior for $\stash$ occurs in the first step of a round, and the above formula
actually gives an upper bound on the amount of keys in the $\stash$.

Ignoring smaller terms as $s_0$ grows and $\epsilon$ goes to $0$, we can rewrite
the expression as$ n \frac{1-\epsilon}{4s_0}{(1-s_0\epsilon)}^2 $.

In particular, when $\epsilon = 0$, we expect the additional stack size (wrt.
a uniform hash function) to grow as $\frac n {4s_0}$.

\section{Keeping an external-memory copy of the stash}%
\label{sec:ext-stash}

In practical applications, it may be useful to keep a copy of the $\stash$
on external memory (for example, to have a copy in case of application crashes).
We now show a variation on our hash table that achieves this at the cost
of increasing the number of block transfer for the updates by an
expected constant factor.
To achieve this, we treat the ``leftover'' space in underflown buckets
as an array (because of our allocation rule, we know that there is always
enough ``leftover'' space to store the full contents of the $\stash$). We
start by filling it from the lowest-indexed bucket and proceed on towards
buckets with bigger indexes.

To insert a key in the $\stash$ on external memory, we try to fit it in the last underflown
block used. If the block is full, we find a new block by scanning all the
buckets on the left until we find a non-full one: this process will terminate
in $O(1)$ block transfers w.h.p.\footnote{If we want to reduce the
  number of block transfers, we can
keep an in-memory dictionary that holds the indexes of the buckets that are
full,  because of keys not in the $\stash$, and perform our scans on that data
structure: the expected number of full buckets is smaller than the expected
stash size.}, as a bucket is not full with probability at least $\frac 12$.

Deletion works in a similar way: if the key to be deleted is not in the last
position of our virtual array, we swap it with the last one and go back one
position.

If we want to insert a ``legitimate'' key in a block that has stash keys in
it, we can identify a stash key since it has the incorrect hash for its
bucket. We can then move it to the front of the virtual array and proceed as
usual. We proceed in the opposite way if space is freed up in a block that
should be used in the virtual array.

It remains to see how we reassign the keys after a
$\newBucket()$ or $\freeBucket()$ operation. In the worst case, it may
cause us to perform $O(s_0B)$ stash operations, which could require up to
$O(s_0B)$ block transfers. However, we may decide to delay those stash operations, doing
$O(s_0)$ of them on each of the subsequent update operations, without causing
significant changes in how the hash table performs.

We can compute the expected number of stash operations and prove that
this method performs better in expectation, without requiring $O(s_0)$
operations per update. We consider the case of $\newBucket()$, as
$\freeBucket()$ behaves in the same way. We use the results of
Subsection~\ref{sub:stashunif}. Since, before looking into the the $\stash$, we
move values within the buckets themselves, the number of $\stash$ operations
may be bounded by the number of empty cells in the buckets after internal
re-arranging. As the load factor on those buckets will be approximately
$1-\epsilon-\frac{1}{s_0}$, we expect $s_0$ of those buckets to still have
space for $B(\epsilon s_0 + 1)$ elements, plus the elements that were put into
the $\stash$.  We can give an upper bound for the elements that are put into
the $\stash$ during this procedure by increasing the load factor to $1-\frac{1}{s_0}$
and using the formulas we obtain from the analysis:
\[
  (s_0+1) \frac{\sqrt{B\left(1-\frac 1{s_0}\right)} e^{-\frac{B}{2s_0^2\left(1-\frac{1}{s_0}\right)}}}{\sqrt{2\pi}}
\]
This can be bounded as $\sqrt{B}+\frac{B}{2\sqrt{\pi}}\tau e^{-\tau^2}$ for some
$\tau$. Since $\tau e^{-\tau^2} < 1$, we can bound the total number of stash
insertions by $O(B)$.

This gives a total of $O(B(1+\epsilon s_0))$ stash operations.
By delaying some of them to the next $O(B)$ updates, we get an expected number
of block transfers per operation of $O(1+\epsilon s_0)$, which is constant as long as $s_0$
is not chosen too big (in particular, it is constant for $s_0 = \frac{2}{\epsilon}$).

\section{Stash sizes for $B=512$ and $B=2048$}

\begin{table*}[h!]
\hspace*{-18pt}
\begin{tabular}{|r|rr|rr|rr|rr|rr|rr|}
  \hline
  \multicolumn{1}{|c|}{\multirow{3}{*}{$s_0$}}  & \multicolumn{12}{c|}{$\epsilon$} \\
  \cline{2-13}
  & \multicolumn{2}{c|}{\small 0} & \multicolumn{2}{c|}{\small 0.001} & \multicolumn{2}{c|}{\small 0.01} & \multicolumn{2}{c|}{\small 0.03} & \multicolumn{2}{c|}{\small 0.05} & \multicolumn{2}{c|}{\small 0.1} \\
  & \multicolumn{1}{c}{real} & \multicolumn{1}{c|}{est.} & \multicolumn{1}{c}{real} & \multicolumn{1}{c|}{est.} & \multicolumn{1}{c}{real} & \multicolumn{1}{c|}{est.} & \multicolumn{1}{c}{real} & \multicolumn{1}{c|}{est.} & \multicolumn{1}{c}{real} & \multicolumn{1}{c|}{est.} & \multicolumn{1}{c}{real} & \multicolumn{1}{c|}{est.} \\
  \hline
  1 & \small 17.2\% & \small 18.9\% & \small 17.1\% & \small 18.8\% & \small 16.7\% & \small 17.9\% & \small 15.9\% & \small 16.1\% & \small 15.1\% & \small 14.7\% & \small 13\% & \small 12\% \\
  4 & \small 5.6\% & \small 7.3\% & \small 5.5\% & \small 7.2\% & \small 5.1\% & \small 6.4\% & \small 4.2\% & \small 4.8\% & \small 3.4\% & \small 3.6\% & \small 1.7\% & \small 1.7\% \\
  16 & \small 2.2\% & \small 3.3\% & \small 2.1\% & \small 3.2\% & \small 1.7\% & \small 2.4\% & \small 1\% & \small 1\% & \small 0.5\% & \small 0.3\% & \bf \small 0.06\% & \bf \small 0.4\% \\
  32 & \small 1.9\% & \small 2.5\% & \small 1.8\% & \small 2.4\% & \small 1.4\% & \small 1.7\% & \small 0.8\% & \small 0.7\% & \bf \small 0.4\% & \bf \small 0.9\% & \bf \small 0.03\% & \bf \small 0.09\% \\
  64 & \small 1.9\% & \small 2.2\% & \small 1.7\% & \small 2.1\% & \small 1.4\% & \small 1.4\% & \bf \small 0.7\% & \bf \small 1.1\% & \bf \small 0.4\% & \bf \small 0.5\% & \bf \small 0.03\% & \bf \small 0.04\% \\
  256 & \small 1.9\% & \small 1.9\% & \small 1.7\% & \small 1.8\% & \bf \small 1.3\% & \bf \small 1.5\% & \bf \small 0.7\% & \bf \small 0.8\% & \bf \small 0.4\% & \bf \small 0.3\% & \bf \small 0.03\% & \bf \small 0.02\% \\
\hline
  ideal & \multicolumn{1}{c}{-} & \small 1.8\% & \multicolumn{1}{c}{-} & \small 1.7\% & \multicolumn{1}{c}{-} & \small 1.3\% & \multicolumn{1}{c}{-} & \small 0.7\% & \multicolumn{1}{c}{-} & \small 0.3\% & \multicolumn{1}{c}{-} & \small 0.01\% \\
  \hline
\end{tabular}
\caption{Percentage of elements on the $\stash$ as $s_0$ and $\epsilon$
change, with $B=512$.}\label{tab:B512}
\end{table*}

\begin{table*}[h!]
\hspace*{-18pt}
\begin{tabular}{|r|rr|rr|rr|rr|rr|rr|}
  \hline
  \multicolumn{1}{|c|}{\multirow{3}{*}{$s_0$}}  & \multicolumn{12}{c|}{$\epsilon$} \\
  \cline{2-13}
  & \multicolumn{2}{c|}{\small 0} & \multicolumn{2}{c|}{\small 0.001} & \multicolumn{2}{c|}{\small 0.01} & \multicolumn{2}{c|}{\small 0.03} & \multicolumn{2}{c|}{\small 0.05} & \multicolumn{2}{c|}{\small 0.1} \\
  & \multicolumn{1}{c}{real} & \multicolumn{1}{c|}{est.} & \multicolumn{1}{c}{real} & \multicolumn{1}{c|}{est.} & \multicolumn{1}{c}{real} & \multicolumn{1}{c|}{est.} & \multicolumn{1}{c}{real} & \multicolumn{1}{c|}{est.} & \multicolumn{1}{c}{real} & \multicolumn{1}{c|}{est.} & \multicolumn{1}{c}{real} & \multicolumn{1}{c|}{est.} \\
  \hline
  1 & \small 17.2\% & \small 18\% & \small 17.1\% & \small 17.9\% & \small 16.7\% & \small 17.1\% & \small 15.9\% & \small 15.6\% & \small 15.1\% & \small 14.4\% & \small 13\% & \small 12\% \\
  4 & \small 5.6\% & \small 6.5\% & \small 5.5\% & \small 6.4\% & \small 5.1\% & \small 5.5\% & \small 4.2\% & \small 4.2\% & \small 3.4\% & \small 3.3\% & \small 1.6\% & \small 1.7\% \\
  16 & \small 1.6\% & \small 2.4\% & \small 1.5\% & \small 2.3\% & \small 1.1\% & \small 1.5\% & \small 0.5\% & \small 0.5\% & \small 0.2\% & \small 0.06\% & \bf \small 0.0009\% & \bf \small 0.02\% \\
  32 & \small 1.1\% & \small 1.7\% & \small 1\% & \small 1.6\% & \small 0.6\% & \small 0.8\% & \small 0.2\% & \small 0.09\% & \bf \small 0.03\% & \bf \small 0.2\% & \bf \small 0.0002\% & \bf \small 0.0002\% \\
  64 & \small 0.9\% & \small 1.3\% & \small 0.8\% & \small 1.2\% & \small 0.5\% & \small 0.5\% & \bf \small 0.1\% & \bf \small 0.3\% & \bf \small 0.02\% & \bf \small 0.05\% & \bf \small 0.0001\% & \bf \small 1e-05\% \\
  256 & \small 0.9\% & \small 1\% & \small 0.8\% & \small 0.9\% & \bf \small 0.5\% & \bf \small 0.6\% & \bf \small 0.1\% & \bf \small 0.1\% & \bf \small 0.01\% & \bf \small 0.01\% & \bf \small 0.0002\% & \bf \small 1e-06\% \\
\hline
  ideal & \multicolumn{1}{c}{-} & \small 0.9\% & \multicolumn{1}{c}{-} & \small 0.8\% & \multicolumn{1}{c}{-} & \small 0.5\% & \multicolumn{1}{c}{-} & \small 0.09\% & \multicolumn{1}{c}{-} & \small 0.008\% & \multicolumn{1}{c}{-} & \small 4e-07\% \\
  \hline
\end{tabular}
\caption{Percentage of elements on the $\stash$ as $s_0$ and $\epsilon$
change, with $B=2048$.}\label{tab:B2048}
\end{table*}

\newpage
\section{Pseudocode}

\begin{algorithm}[h]
\small
\DontPrintSemicolon%
\SetKwProg{myproc}{Function}{}{}
  \myproc{$\lookup(x)$}{
    \lIf{$x \in \stash$}{\Return{$\true$}}
    $b \gets \findBucket(h(x)/|I|)$\;
    \lIf{$x \in \block(b)$}{\Return{$\true$}}
    \Return{$\false$}   
}
\caption{Lookup algorithm}%
\label{alg:lookup}
\end{algorithm}

\begin{algorithm}[h]
\small
\DontPrintSemicolon%
\SetKwProg{myproc}{Function}{}{}
  \myproc{$\ins(x)$}{
    \lIf{$\lookup(x)$}{\Return{$\false$}}
    $n \gets n+1$ \quad //initially $n=0$\;
    \If{$\lceil \frac{n}{B (1-\epsilon)}\rceil > \numBuckets()$}{
      $\distribute( \newBucket() )$\;
    }
    $b \gets \findBucket(h(x)/|I|)$\;
    \If{$|\block(b)| < B$}{
      $\block(b) = \block(b) \cup \{ x \}$\;
    }
    \Else%
    {
      $\stash = \stash \cup \{ x \}$\;
    }
    \Return{$\true$}\;
  }
\caption{Insertion algorithm}%
\label{alg:insert}
\end{algorithm}

\begin{algorithm}[h]
\small
\DontPrintSemicolon%
\SetKwProg{myproc}{Function}{}{}
  \myproc{$\del(x)$}{
    $\found = \false$\;
    \If{$x \in \stash$}{
      $\stash = \stash \setminus \{x\}$\;
      $\found = \true$\;
    }
    \Else%
    {
      $b \gets \findBucket(h(x)/|I|)$\;
      \If{$x \in \block(b)$}{
        $\block(b) = \block(b) \setminus \{ x \}$\;
        $\found = \true$\;
      }
    }
    \If{$\found$}{
      $n \gets n-1$\;
      \If{$n > 0$ and $\lceil \frac{n}{B (1-\epsilon)}\rceil < \numBuckets() - 1$}{
        $\distribute( \freeBucket() )$\;
      }
    }
    \Return{$\found$}
  }
\caption{Deletion algorithm}%
\label{alg:delete}
\end{algorithm}

\end{document}